\documentclass[11pt]{article}

\usepackage[T1]{fontenc}
\usepackage[utf8]{inputenc}
\usepackage{lmodern}
\usepackage[margin=1in]{geometry}
\usepackage{amsmath,amssymb}
\usepackage{graphicx}
\usepackage{booktabs}
\usepackage{array}
\usepackage{xcolor}
\usepackage[numbers,sort&compress]{natbib}
\usepackage{xurl}
\usepackage[colorlinks=true,linkcolor=blue!50!black,citecolor=blue!50!black,urlcolor=blue!50!black]{hyperref}

\graphicspath{{figures/}}
\newcommand{\plainlink}[2]{{\hypersetup{urlcolor=.}\href{#1}{#2}}}
\newcommand{\ci}[2]{[#1, #2]}
\newcommand{\tok}[1]{\texttt{#1}}
\newcommand{\corraddress}{\texttt{soroush.mehri@elementera.com}}

\title{AI Crawler Access for AEO and GEO:\\
How Startup Websites Treat AI Search and\\
AI Training Crawlers in robots.txt}

\author{\plainlink{https://www.elementera.com/team/soroush-mehri?r=zAtpv9}{Soroush Mehri} \qquad \plainlink{https://www.elementera.com/team/mojtaba-ebadi?r=zAtpv9}{Mojtaba Ebadi} \qquad \plainlink{https://www.elementera.com/team/dr-maziar-gomrokchi?r=zAtpv9}{Maziar Gomrokchi} \qquad \plainlink{https://www.elementera.com/team/dr-mehrdad-mehrbod?r=zAtpv9}{Mehrdad Mehrbod}\\[4pt]
\plainlink{https://www.elementera.com/?r=zAtpv9}{Elementera AI}\\[4pt]
\small Equal contribution.\\
\small Correspondence: Soroush Mehri, \corraddress\\
\small Project page: \plainlink{https://www.elementera.com/resources/ai-crawler-access-report?r=zAtpv9}{\texttt{www.elementera.com/resources/ai-crawler-access-report}}}

\date{November 2026}

\begin{document}
\maketitle

\begin{abstract}
AI search crawlers are rarely blocked in the robots.txt files of startup websites, and AI training crawlers are blocked more often, mostly by name. For answer engine optimization (AEO) and generative engine optimization (GEO), robots.txt is the file where a website tells AI crawlers whether they may read it. The major AI vendors now publish separate crawler names for search (for example OAI-SearchBot), for fetching a page at a user's request, and for collecting training data (for example GPTBot). Under a method fixed before data collection, we drew a stratified random sample of 600 websites of Y Combinator-backed B2B, fintech and developer-tool companies and read the robots.txt file of each main host on 7 October 2026, resolving the rule that applies to 22 crawler tokens on six common paths with an RFC 9309 parser. Of the 569 sites that gave a readable answer, 102 (17.9 percent; 95 percent interval 15.0 to 21.3) name at least one AI crawler and 467 (82.1 percent; 78.7 to 85.0) name none. Three of the 569 (0.5 percent; 0.2 to 1.5) block an AI search crawler, all three in the ambiguous category of our locked rules, and no site blocks an AI search crawler while leaving every AI training crawler open (0 of 569; 0.0 to 0.7). Eighteen of the 569 (3.2 percent; 2.0 to 4.9) block at least one AI training crawler, and in 15 of those 18 the file names the blocked crawler. Fifteen of the 569 (2.6 percent; 1.6 to 4.3) block an AI training crawler while leaving every AI search crawler open. robots.txt states a request, not an access control, and we do not measure crawler behaviour. The results describe young, venture-backed companies and do not generalize to publishers or to the web at large. We release the per-site dataset without site identities.
\end{abstract}

\section{Introduction}
\label{sec:intro}

Generative AI systems reach the open web in at least three ways. They collect pages to build training corpora; they index pages so that an answer engine can retrieve and cite them in response to a query; and they fetch a single page when a person asks an assistant to read it. The large vendors now publish a separate crawler name, or user-agent token, for each of these roles. OpenAI documents \tok{GPTBot} for training, \tok{OAI-SearchBot} for search and \tok{ChatGPT-User} for user requests \citep{openaibots}; Anthropic documents \tok{ClaudeBot}, \tok{Claude-SearchBot} and \tok{Claude-User} \citep{anthropiccrawl}; Perplexity documents \tok{PerplexityBot} and \tok{Perplexity-User} \citep{perplexitybots}; and Google documents \tok{Google-Extended}, a control token that governs some uses of content by its generative models without changing what Googlebot crawls \citep{googlecrawlers}. A site operator who wants to appear in AI search answers but not in training data can, in principle, say so in robots.txt by allowing one token and disallowing another. For answer engine optimization (AEO) and generative engine optimization (GEO), the work of making a site visible in AI search answers, robots.txt is the file where a site states which AI crawlers may read it.

\paragraph{In brief.} We read the robots.txt files of 569 startup websites on 7 October 2026. Most of them, 467 of 569, do not mention any AI crawler. Only 3 of 569 block an AI search crawler, and all three are ambiguous cases under our locked rules. AI training crawlers are blocked more often, on 18 of 569 sites, and in 15 of those 18 the file names the crawler it blocks. robots.txt is a request to crawlers, not a lock, and we do not measure whether crawlers obey it.

robots.txt is a request, not an access control. RFC 9309 states that its rules ``are not a form of access authorization'' \citep{rfc9309}, and a controlled study found that compliance with robots.txt directives varies by crawler category and falls as directives become stricter \citep{kim2025scrapers}. A content-delivery network or firewall can refuse a crawler that robots.txt allows, and the reverse. What robots.txt does record is a stated preference, written by the operator or by a platform on the operator's behalf, and this is what we measure.

Most measurement of AI-crawler rules has studied news publishers and large, popular sites \citep{fletcher2024news,longpre2024consent,liu2025somesite,steinacker2025misinformation}. Those sites have content that is valuable as training data and, in many cases, a commercial reason to restrict it. Young software companies are in a different position: their websites exist to be found, and their content is mostly marketing and documentation. Whether such sites distinguish search crawlers from training crawlers, and whether any block they carry was written on purpose, has not, to our knowledge, been measured with a preregistered method on a probability sample.

\paragraph{Contributions.}
\begin{enumerate}
\item A probability sample of 600 company websites drawn from a pinned public directory, with a method for reading robots.txt and for labelling each block that was fixed before the sampled sites were fetched (Section~\ref{sec:method}).
\item Site-level estimates, each with its numerator, denominator and a 95\% Wilson interval, of how often these sites name AI crawlers and block AI search, AI user-request and AI training crawlers, and of how each block arises (Section~\ref{sec:results}).
\item A descriptive count of \tok{Content-Signal} lines in the same files, and a released per-site dataset without site identities (Section~\ref{sec:repro}).
\end{enumerate}

\paragraph{What this paper does not claim.} We do not measure whether any crawler obeys these files, whether a site is reachable by a crawler, or whether being crawled affects how often a site appears in AI answers. Labels such as ``named'' describe the text of a file, not the intent of the company behind it. The authors run a company that sells measurement of visibility in AI answer engines; the conflict of interest is stated at the end of the paper.

\section{Related work}
\label{sec:related}

\paragraph{robots.txt as a measurement object.} Early large-scale studies counted the adoption of robots.txt and the crawlers it names. Sun, Zhuang and Giles examined 7{,}593 websites and found that most files with rules address all crawlers through the \tok{*} group \citep{sun2007robots}; Kolay et al.\ extended the analysis to a much larger crawl and to differences in how files treat individual search engines \citep{kolay2008robots}. RFC 9309 later standardized the parsing rules that we apply, including group matching, longest-match precedence and the handling of fetch errors \citep{rfc9309}.

\paragraph{AI crawlers in robots.txt.} Fletcher measured, from archived files, how many of the most-used news websites in ten countries blocked AI crawlers at the end of 2023, and found that a large share of them blocked OpenAI's crawler \citep{fletcher2024news}. Longpre et al.\ audited the robots.txt files and terms of service of the web domains behind widely used training corpora and documented a rapid rise in restrictions over 2023 and 2024, with OpenAI's crawlers restricted more often than those of other developers \citep{longpre2024consent}. Liu et al.\ combined a survey of artists with a longitudinal measurement of AI-crawler directives in the robots.txt files of popular sites, found an initial surge in such directives followed by slower growth, and reported a small but growing number of sites that explicitly invite AI crawlers \citep{liu2025somesite}. Steinacker-Olsztyn, Gosain and Dao compared reputable and misinformation news sites and found that the former restrict AI agents far more often \citep{steinacker2025misinformation}. Kim et al.\ measured, with controlled robots.txt files on their own servers, how often self-declared bots fetch and obey robots.txt, and found that compliance varies by bot category \citep{kim2025scrapers}.

\paragraph{How this study differs.} The studies above target publishers, popular domains or corpus sources, and most treat ``AI crawler'' as one class or focus on training crawlers. We study a different population, young B2B software companies, with a probability sample from a pinned frame; we resolve each crawler token separately and group the tokens by the role the vendor documents (search, user request, training); and we label each block by how it arises under rules written before data collection. Our numbers are not comparable to publisher-focused estimates and are not intended to be.

\paragraph{A newer signal.} Cloudflare's Content Signals Policy adds a \tok{Content-Signal} line to robots.txt that states preferences for search, AI input and AI training, and Cloudflare announced that it would add such a line to the robots.txt files it manages for its customers \citep{cloudflare2025signals}. We count such lines in our sample but do not evaluate whether any system reads them.

\section{Data and method}
\label{sec:method}

\subsection{Preregistration}

The method document was written and its SHA-256 hash recorded on 6 October 2026, before any sampled site's robots.txt was fetched. The sections fixing the parsing rules and the block categories (Sections~\ref{sec:parse} and~\ref{sec:cats} below) were marked as locked. The preregistration is internal: the document and its hash are kept in the authors' repository and were not filed with a public registry. The first recorded hash was of an uncommitted version of the document and cannot be checked by a third party; the first verifiable lock is the committed version of the document, whose hash is recorded alongside it. By the time of that commit, three sampled sites had been fetched in a labelled pilot, and no locked rule changed after the pilot. One clarification was recorded before any sampled site was fetched: the list of training-crawler tokens used in category rule~5 (Section~\ref{sec:cats}) is the set of tokens with class ``training'' in the registry plus the legacy token \tok{anthropic-ai}.

\subsection{Sampling frame and sample}

The frame is the public Y Combinator company directory as mirrored in the \tok{yc-oss/api} repository, file \tok{companies/all.json}, pinned to a single commit \citep{ycoss}; the file's checksum is recorded. Of its 6{,}273 records we kept companies whose status is \emph{Active} or \emph{Public} (1{,}915 excluded), whose industry is B2B or Fintech (1{,}599 excluded), that list a website whose host is not an IP address (4 excluded), and whose registrable domain, computed with a pinned copy of the Public Suffix List \citep{psl}, is not a shared platform such as a code-hosting or social-network domain (3 excluded). After deduplication on registrable domain the frame holds 2{,}752 sites.

We stratified the frame into 12 strata: three industry groups (developer tools and infrastructure; fintech; other B2B) by four batch-year bands (2019 or earlier; 2020--2022; 2023--2024; 2025 or later). We allocated 600 sites to strata in proportion to stratum size and drew them without replacement within each stratum with a fixed seed. Proportional allocation makes the sample self-weighting, so unweighted proportions estimate frame proportions. The sample holds 313 other-B2B, 180 developer-tool and 107 fintech sites before fetch outcomes are applied.

\subsection{Fetching}

For each sampled site we requested only \tok{https://<host>/robots.txt}, once, with a user-agent string that identifies the research and links to a description of it, never with a crawler's user agent, and without credentials. Connect and total timeouts were 8 and 15 seconds. Redirects were followed by hand for at most five hops and only within the same registrable domain; a redirect to another registrable domain was recorded as undetermined. One host was not contacted because our network policy did not permit it. The baseline reading ran on 7 October 2026 between 03:01:34 and 03:02:15 UTC.

\subsection{Reading robots.txt}
\label{sec:parse}

Parsing follows RFC 9309 \citep{rfc9309}: the group whose user-agent line best matches a crawler applies, and if any named group matches, the \tok{*} group is ignored for that crawler; within a group the longest matching rule wins and \tok{Allow} wins a tie; \tok{*} and \tok{\$} wildcards are supported; at most 500\,KiB of the body is read. The fetch outcome maps to an access state as RFC 9309 prescribes. A 2xx response is parsed. A 4xx response (including 401, 403 and 429) means no restrictions and is counted as ``no robots.txt'', all crawlers allowed; the 401, 403 and 429 cases are also counted separately because they may be a firewall refusing our fetcher. A 5xx response, timeout, DNS or TLS failure would mean complete disallow under the RFC; we report such sites as unreachable and never count them as blocks. Off-domain redirects and redirect loops are undetermined.

For each parsed file and each of 22 crawler tokens we evaluated six probe paths: \tok{/}, \tok{/pricing}, \tok{/blog/}, \tok{/docs/}, \tok{/about} and \tok{/product}. A crawler is \emph{blocked} on a site if at least one probe path is disallowed for it (on all six paths, or on some) and \emph{allowed} otherwise; disallow rules for other paths (for example \tok{/admin/}) are not blocks. We record the decision source: an explicit group that names the crawler, the wildcard group only, no applicable group, or no robots.txt. We say a file \emph{names} a crawler when the governing group for that crawler names it.

\subsection{Crawler registry}

Table~\ref{tab:registry} lists the tokens. Classes follow the vendors' documentation where we could read it \citep{openaibots,anthropiccrawl,perplexitybots,googlecrawlers} and were cross-checked against a community registry pinned to a commit \citep{airobotstxt}. The fourteen \emph{AI crawlers} of the headline are the search, user-request and training tokens.\footnote{The authors keep a list of AI crawlers and their documented roles (\plainlink{https://www.elementera.com/ai-crawlers?r=zAtpv9}{AI crawlers list}), also as JSON (\plainlink{https://www.elementera.com/tools/ai-crawler-registry?r=zAtpv9}{AI crawler registry}). It has been updated since this study and may differ from Table~\ref{tab:registry}.} Googlebot, Bingbot and Applebot serve as web-search references. Five further tokens (\tok{OAI-AdsBot}, \tok{Google-CloudVertexBot}, \tok{GoogleOther}, \tok{Amazonbot} and the legacy \tok{anthropic-ai}) are resolved but not reported in the headline.

\begin{table}[t]
\centering
\small
\caption{Crawler tokens resolved on every file, by the role the vendor documents.}
\label{tab:registry}
\begin{tabular}{@{}l>{\raggedright\arraybackslash}p{7.4cm}>{\raggedright\arraybackslash}p{4.6cm}@{}}
\toprule
Role & Tokens & Vendors \\
\midrule
AI search & \tok{OAI-SearchBot}, \tok{Claude-SearchBot}, \tok{PerplexityBot} & OpenAI, Anthropic, Perplexity \\
AI user request & \tok{ChatGPT-User}, \tok{Claude-User}, \tok{Perplexity-User}, \tok{Meta-ExternalFetcher} & OpenAI, Anthropic, Perplexity, Meta \\
AI training & \tok{GPTBot}, \tok{ClaudeBot}, \tok{Google-Extended}, \tok{CCBot}, \tok{Applebot-Extended}, \tok{Meta-ExternalAgent}, \tok{Bytespider} & OpenAI, Anthropic, Google, Common Crawl, Apple, Meta, ByteDance \\
Web search reference & \tok{Googlebot}, \tok{Bingbot}, \tok{Applebot} & Google, Microsoft, Apple \\
\bottomrule
\end{tabular}
\end{table}

\subsection{How a block arises}
\label{sec:cats}

Every blocked site--crawler pair receives exactly one category, by the first of the following rules that matches. The rules were written before any sampled site was fetched.
\begin{enumerate}
\item \emph{Likely accidental (contradiction)}: a search or user-request crawler is blocked although the same file carries a \tok{Content-Signal} line stating \tok{search=yes}.
\item \emph{Platform default}: the deciding rule lies between the markers that delimit a Cloudflare-managed section of the file. A platform wrote the rule; the operator may or may not have chosen it.
\item \emph{Named}: the decision source is an explicit group that names the crawler.
\item \emph{Blanket}: the decision source is the wildcard group and Googlebot and Bingbot are also blocked on all six paths.
\item \emph{Likely accidental (training rule caught search)}: a search or user-request crawler is blocked by the wildcard group and the file has a named group made only of training crawlers that disallows the same path.
\item \emph{Ambiguous}: every other block, for example a wildcard disallow that applies to every crawler not named elsewhere in the file.
\end{enumerate}
``Named'' means that someone, or some template, wrote the crawler's name; it does not mean that the company weighed the decision.

\subsection{Statistics}

The site is the unit throughout, because the rules for different crawlers in one file are not independent. Every proportion is reported as $k$ of $n$ with a 95\% Wilson score interval \citep{wilson1927,brown2001interval},
\begin{equation}
\frac{\hat p + \frac{z^2}{2n}}{1 + \frac{z^2}{n}} \pm \frac{z}{1 + \frac{z^2}{n}} \sqrt{\frac{\hat p (1-\hat p)}{n} + \frac{z^2}{4n^2}}, \qquad z = 1.96,
\end{equation}
which behaves well for the small counts that dominate our results. The denominator for per-crawler and site-level figures is the set of sites with a determined result (a parsed file or no file); unreachable, undetermined and unfetched sites are reported as their own counts. The intervals ignore both the stratification and the finite-population correction (the sample is about 22\% of the frame); both omissions make them slightly conservative. Following the locked method, we claim a difference between two proportions only when their intervals do not overlap, and we report overlap as ``not distinguishable at this $n$''. This rule is conservative.

\section{Results}
\label{sec:results}

\subsection{Fetch outcomes}

Table~\ref{tab:fetch} gives the outcome for all 600 sampled sites. 496 returned a file that we parsed and 73 had no robots.txt, so 569 sites have a determined result. Eleven parsed responses were HTML pages served at the robots.txt path; they contain no groups, so every crawler is allowed on them. Seven sites answered our fetcher with 401, 403 or 429, which RFC 9309 treats as no restrictions; those responses may come from a firewall and are a known source of undercounting (Section~\ref{sec:limits}).

\begin{table}[t]
\centering
\small
\caption{Fetch outcomes at baseline (7 October 2026). Intervals are 95\% Wilson.}
\label{tab:fetch}
\begin{tabular}{@{}lrrrl@{}}
\toprule
Outcome & $k$ & $n$ & \% & 95\% interval \\
\midrule
robots.txt parsed (2xx) & 496 & 600 & 82.7 & \ci{79.4}{85.5} \\
No robots.txt (4xx) & 73 & 600 & 12.2 & \ci{9.8}{15.0} \\
\quad of which 401, 403 or 429 & 7 & 600 & 1.2 & \ci{0.6}{2.4} \\
Unreachable (5xx, timeout, DNS, TLS) & 11 & 600 & 1.8 & \ci{1.0}{3.3} \\
Undetermined (off-domain redirect, loop) & 19 & 600 & 3.2 & \ci{2.0}{4.9} \\
Not fetched (network policy) & 1 & 600 & 0.2 & \ci{0.0}{0.9} \\
\midrule
HTML page served as robots.txt & 11 & 496 & 2.2 & \ci{1.2}{3.9} \\
\bottomrule
\end{tabular}
\end{table}

\subsection{Most files do not mention AI crawlers}

Of the 569 readable sites, 102 have a group that names at least one of the fourteen AI crawlers (17.9\%, \ci{15.0}{21.3}); the other 467 (82.1\%, \ci{78.7}{85.0}) name none, so every AI crawler is allowed on them by default or governed only by the wildcard group. The most often named tokens are \tok{GPTBot} (98 files), \tok{ChatGPT-User} and \tok{PerplexityBot} (93 each), \tok{ClaudeBot} (86), \tok{Google-Extended} (85) and \tok{OAI-SearchBot} (81) (Figure~\ref{fig:crawler}b).

\subsection{AI search crawlers are rarely blocked}

Three of 569 sites block at least one of \tok{OAI-SearchBot}, \tok{Claude-SearchBot} and \tok{PerplexityBot} (0.5\%, \ci{0.2}{1.5}); two block all three (0.4\%, \ci{0.1}{1.3}). All three sites fall in the ``ambiguous'' category of Section~\ref{sec:cats}. No site blocks an AI search crawler while leaving every training crawler open (0 of 569, \ci{0.0}{0.7}). No file carries a blanket block (0 of 569, \ci{0.0}{0.7}), and neither ``likely accidental'' rule fired.

User-request fetchers are blocked on 5 of 569 sites (0.9\%, \ci{0.4}{2.0}). A named group blocks \tok{ChatGPT-User} on one site and \tok{Meta-ExternalFetcher} on one site.

\begin{figure}[t]
\centering
\includegraphics[width=\linewidth]{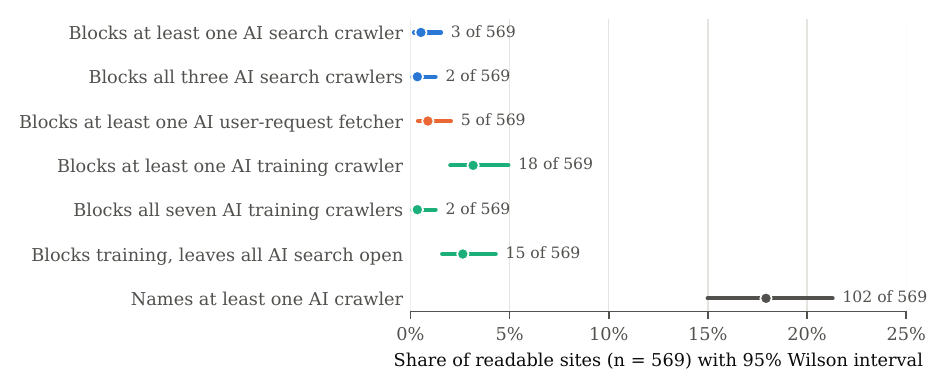}
\caption{Site-level shares at baseline, 7 October 2026, among the 569 sites with a determined result. Dots are point estimates, bars 95\% Wilson intervals, labels the counts. Search, user-request and training crawlers are the groups of Table~\ref{tab:registry}.}
\label{fig:headline}
\end{figure}

\subsection{AI training crawlers are blocked more often, mostly by name}

\begin{figure}[t]
\centering
\includegraphics[width=\linewidth]{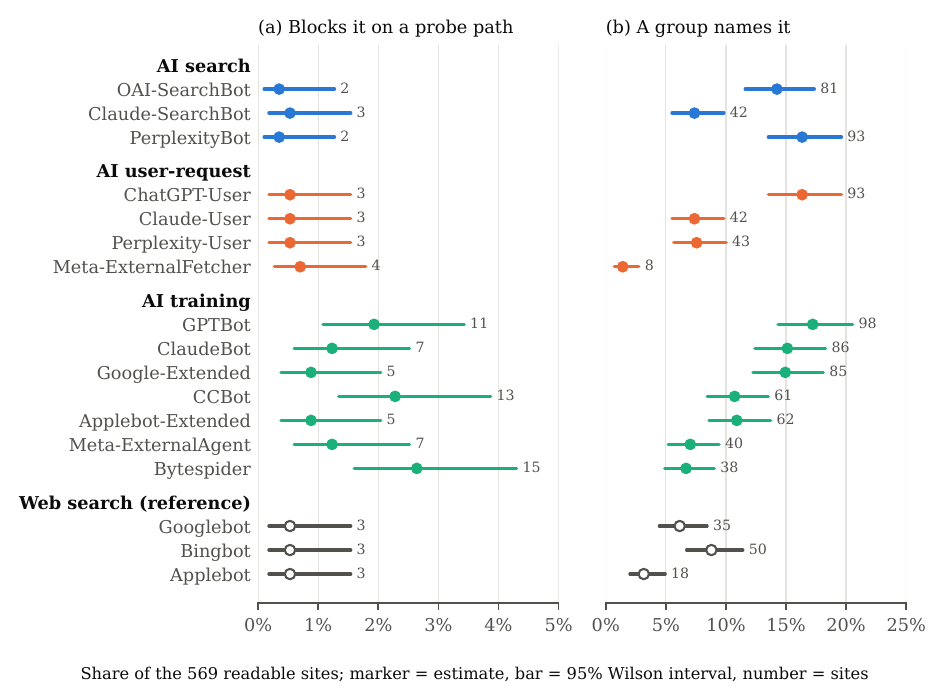}
\caption{Per crawler, at baseline. (a) Share of the 569 readable sites that block the crawler on at least one of the six probe paths. (b) Share whose governing group names the crawler. Hollow markers are the web-search references. Dots are estimates, bars 95\% Wilson intervals, numbers the counts of sites.}
\label{fig:crawler}
\end{figure}

Eighteen of 569 sites block at least one AI training crawler (3.2\%, \ci{2.0}{4.9}); two block all seven (0.4\%, \ci{0.1}{1.3}). The intervals for AI search blocks and AI training blocks do not overlap (Figure~\ref{fig:headline}). In 15 of the 18 sites (83.3\%, \ci{60.8}{94.2}) the file names the blocked training crawler. Fifteen of 569 sites block at least one training crawler while leaving all three AI search crawlers open (2.6\%, \ci{1.6}{4.3}), which is the separation the vendors' token design permits. The tokens blocked most often are \tok{Bytespider} (15 sites), \tok{CCBot} (13) and \tok{GPTBot} (11) (Figure~\ref{fig:crawler}a). At baseline no training block lay inside a platform-managed section; one of the 496 parsed files contained the Cloudflare markers at all.

\subsection{Industry groups}

Table~\ref{tab:industry} splits the two headline figures by industry group. The intervals overlap in every comparison, so no difference between groups is distinguishable at these sample sizes.

\begin{table}[t]
\centering
\small
\caption{Sites blocking at least one AI search or AI training crawler, by industry group (baseline). Intervals are 95\% Wilson.}
\label{tab:industry}
\begin{tabular}{@{}lrrlrl@{}}
\toprule
Industry group & $n$ & AI search & 95\% interval & AI training & 95\% interval \\
\midrule
Other B2B & 298 & 2 (0.7\%) & \ci{0.2}{2.4} & 10 (3.4\%) & \ci{1.8}{6.1} \\
Developer tools and infrastructure & 168 & 1 (0.6\%) & \ci{0.1}{3.3} & 3 (1.8\%) & \ci{0.6}{5.1} \\
Fintech & 103 & 0 (0.0\%) & \ci{0.0}{3.6} & 5 (4.9\%) & \ci{2.1}{10.9} \\
\midrule
All & 569 & 3 (0.5\%) & \ci{0.2}{1.5} & 18 (3.2\%) & \ci{2.0}{4.9} \\
\bottomrule
\end{tabular}
\end{table}

\subsection{Content-Signal lines}

Twenty-six of 569 sites carry a \tok{Content-Signal} line in robots.txt (4.6\%, \ci{3.1}{6.6}; 26 of the 496 parsed files). This count was not part of the locked analysis plan and is descriptive. A \tok{Content-Signal} line states a preference about use after access; it does not change what any path rule allows, and our classification uses it only in category rule~1.

\section{Limitations}
\label{sec:limits}

\paragraph{One frame.} The sample represents B2B and fintech companies listed as active or public in one directory on one date. These companies are young, venture-funded and mostly English-language, and many sites are months old and built on templates. The results do not describe publishers, e-commerce, the web at large, or companies outside this directory. Frames for news publishers and for Canadian companies in the same directory have been built for future readings; neither is analysed here.

\paragraph{robots.txt only.} We read a stated request. We do not observe whether crawlers fetch or obey robots.txt, which varies by crawler \citep{kim2025scrapers}, nor blocks enforced at a content-delivery network or firewall, which can exist whatever robots.txt says. The seven sites that answered our fetcher with 401, 403 or 429 are counted as unrestricted under RFC 9309, which may hide network-level refusals.

\paragraph{Main host and six paths.} We read the robots.txt of the host listed in the directory, after same-domain redirects. Documentation, blog and application subdomains have their own files and were not read. A block on a path outside the six probe paths is not counted.

\paragraph{Parser details.} The parser matches a user-agent line as a prefix of the crawler token, so a group for \tok{Applebot} also governs \tok{Applebot-Extended} when no group names the latter; this applied to five sites. Strict RFC 9309 matching compares the product token and differs from prefix matching only for unusual user-agent values.

\paragraph{Classes and labels.} Classes follow vendor documentation where we could read it; for Apple, Meta, Common Crawl, Microsoft and Amazon the class was taken from the cited registry and the vendor's page was not re-read, and \tok{Bytespider} has no vendor documentation. The block categories are inferences from file text. We did not contact any company.

\paragraph{Time.} The results come from one reading on one day. robots.txt files change, and the figures describe the files as they were served on that day.

\paragraph{Exploratory parts.} The \tok{Content-Signal} count was not preregistered.

\section{Ethics}

The reading sent one GET request for \tok{/robots.txt} per site, with a user agent that identifies the research and links to a description of it. No other path was requested, no crawler was impersonated, and no credentials were used. No site, company or domain is named in this paper or in the released data; each site carries a random identifier.

\section{Reproducibility statement}
\label{sec:repro}

The sampling frame is rebuilt from the two pinned public sources by a deterministic script. The per-site dataset of the baseline (600 rows; for each site its random identifier, industry group, batch band, fetch outcome and, for each of the 17 headline and reference tokens, its state, decision source and category) recomputes every figure in Section~\ref{sec:results} except the \tok{Content-Signal} count, which depends on file text. The per-site dataset and the method are published on the project's page on elementera.com, \url{https://www.elementera.com/resources/ai-crawler-access-report}, which links to the dataset download. For this paper the classification and analysis of the baseline were re-run from the archived raw responses and reproduce the summary and the dataset exactly. The figures are produced from these outputs by a script that recounts each plotted figure from the per-site dataset and checks it against the summary. The raw robots.txt bodies are not released, because they identify the companies.

\section{Conclusion}

On a probability sample of 600 young B2B software and fintech companies, AI crawlers are rarely blocked in robots.txt. Most files, 467 of 569, name no AI crawler at all. AI search crawlers are blocked on only 3 of 569 readable sites. All three are in the ambiguous category of our locked rules, and no site blocks a search crawler while leaving every training crawler open. AI training crawlers are blocked more often, on 18 of 569 sites, and in 15 of those 18 the file names the blocked crawler. On 15 of 569 sites a training crawler is blocked while every AI search crawler stays open, which is the separation the vendors' tokens are designed to express. For this population, robots.txt mostly records silence towards AI crawlers. For AEO and GEO, the practical reading is that almost none of these sites ask AI search crawlers to stay away. Measurements of AI-crawler policy should distinguish search from training tokens and state that robots.txt is a request.\footnote{Site owners can see which AI crawlers their own robots.txt blocks with the \plainlink{https://www.elementera.com/tools/ai-crawler-check?r=zAtpv9}{AI crawler checker}, draft rules with the \plainlink{https://www.elementera.com/tools/robots-txt-ai-generator?r=zAtpv9}{robots.txt generator for AI crawlers}, and review other technical settings with the \plainlink{https://www.elementera.com/tools/ai-search-technical-checklist?r=zAtpv9}{AI search technical checklist}.}

\section*{Conflict of interest}

All four authors are founders of \plainlink{https://www.elementera.com/?r=zAtpv9}{Elementera AI Inc.}, whose products are \plainlink{https://www.elementera.com/product/lenz?r=zAtpv9}{Lenz}, which measures the visibility of brands and sources in AI answer engines, and \plainlink{https://www.elementera.com/product/originum?r=zAtpv9}{Originum}, which turns what experts know into content and gives each draft an originality score; the company also sells \plainlink{https://www.elementera.com/managed-aeo-services?r=zAtpv9}{managed services} built on both. The founders' earlier peer-reviewed research is listed on the \plainlink{https://www.elementera.com/research-papers?r=zAtpv9}{research papers} page. A site's robots.txt rules for AI crawlers bear on that visibility, and an earlier version of the baseline results appeared as a report on the company's website \citep{elementera2026report}. This work was funded by Elementera. No external grant supported it.

\bibliographystyle{plainnat}
\bibliography{refs}

\begin{thebibliography}{19}
\providecommand{\natexlab}[1]{#1}
\providecommand{\url}[1]{\texttt{#1}}
\expandafter\ifx\csname urlstyle\endcsname\relax
  \providecommand{\doi}[1]{doi: #1}\else
  \providecommand{\doi}{doi: \begingroup \urlstyle{rm}\Url}\fi

\bibitem[{ai.robots.txt contributors}(2026)]{airobotstxt}
{ai.robots.txt contributors}.
\newblock ai.robots.txt.
\newblock \url{https://github.com/ai-robots-txt/ai.robots.txt}, file
  \texttt{robots.json} at commit 9ad8a47e, 2026.

\bibitem[{Anthropic}(2026)]{anthropiccrawl}
{Anthropic}.
\newblock {Anthropic} crawler documentation (help centre article 8896518).
\newblock
  \url{https://support.claude.com/en/articles/8896518-does-anthropic-crawl-data-from-the-web-and-how-can-site-owners-block-the-crawler},
  2026.
\newblock Accessed 6 October 2026.

\bibitem[Brown et~al.(2001)Brown, Cai, and DasGupta]{brown2001interval}
Lawrence~D. Brown, T.~Tony Cai, and Anirban DasGupta.
\newblock Interval estimation for a binomial proportion.
\newblock \emph{Statistical Science}, 16\penalty0 (2):\penalty0 101--133, 2001.
\newblock \doi{10.1214/ss/1009213286}.

\bibitem[{Cloudflare}(2025)]{cloudflare2025signals}
{Cloudflare}.
\newblock Giving users choice with {Cloudflare}'s new {Content Signals Policy}.
\newblock Cloudflare blog,
  \url{https://blog.cloudflare.com/content-signals-policy/}, September 2025.

\bibitem[{Elementera}(2026)]{elementera2026report}
{Elementera}.
\newblock The {AI} crawler access report.
\newblock Elementera,
  \url{https://www.elementera.com/resources/ai-crawler-access-report}, October
  2026.

\bibitem[Fletcher(2024)]{fletcher2024news}
Richard Fletcher.
\newblock How many news websites block {AI} crawlers?
\newblock Factsheet, Reuters Institute for the Study of Journalism, University
  of Oxford, February 2024.

\bibitem[{Google}(2026)]{googlecrawlers}
{Google}.
\newblock {Google} common crawlers documentation.
\newblock
  \url{https://developers.google.com/crawling/docs/crawlers-fetchers/google-common-crawlers},
  2026.
\newblock Accessed 6 October 2026.

\bibitem[Kim et~al.(2025)Kim, Bock, Luo, Liswood, Poroslay, and
  Wenger]{kim2025scrapers}
Taein Kim, Karstan Bock, Claire Luo, Amanda Liswood, Chloe Poroslay, and Emily
  Wenger.
\newblock Scrapers selectively respect robots.txt directives: Evidence from a
  large-scale empirical study.
\newblock In \emph{Proceedings of the 2025 ACM Internet Measurement Conference
  (IMC '25)}, 2025.
\newblock \doi{10.1145/3730567.3764471}.
\newblock arXiv:2505.21733.

\bibitem[Kolay et~al.(2008)Kolay, D'Alberto, Dasdan, and
  Bhattacharjee]{kolay2008robots}
Santanu Kolay, Paolo D'Alberto, Ali Dasdan, and Arnab Bhattacharjee.
\newblock A larger scale study of robots.txt.
\newblock In \emph{Proceedings of the 17th International Conference on World
  Wide Web (WWW '08)}, pages 1171--1172, 2008.
\newblock \doi{10.1145/1367497.1367711}.

\bibitem[Koster et~al.(2022)Koster, Illyes, Zeller, and Sassman]{rfc9309}
Martijn Koster, Gary Illyes, Henner Zeller, and Lizzi Sassman.
\newblock Robots exclusion protocol.
\newblock RFC 9309, Internet Engineering Task Force, September 2022.

\bibitem[Liu et~al.(2025)Liu, Luo, Shan, Voelker, Zhao, and
  Savage]{liu2025somesite}
Enze Liu, Elisa Luo, Shawn Shan, Geoffrey~M. Voelker, Ben~Y. Zhao, and Stefan
  Savage.
\newblock Somesite {I} used to crawl: Awareness, agency and efficacy in
  protecting content creators from {AI} crawlers.
\newblock In \emph{Proceedings of the 2025 ACM Internet Measurement Conference
  (IMC '25)}, 2025.
\newblock \doi{10.1145/3730567.3732913}.
\newblock arXiv:2411.15091.

\bibitem[Longpre et~al.(2024)Longpre, Mahari, Lee, Lund,
  et~al.]{longpre2024consent}
Shayne Longpre, Robert Mahari, Ariel Lee, Campbell Lund, et~al.
\newblock Consent in crisis: The rapid decline of the {AI} data commons.
\newblock In \emph{Advances in Neural Information Processing Systems 37,
  Datasets and Benchmarks Track}, 2024.
\newblock \doi{10.52202/079017-3431}.
\newblock arXiv:2407.14933.

\bibitem[{OpenAI}(2026)]{openaibots}
{OpenAI}.
\newblock {OpenAI} crawler documentation.
\newblock \url{https://developers.openai.com/api/docs/bots}, 2026.
\newblock Accessed 6 October 2026.

\bibitem[{Perplexity}(2026)]{perplexitybots}
{Perplexity}.
\newblock {Perplexity} crawler documentation.
\newblock \url{https://docs.perplexity.ai/docs/resources/perplexity-crawlers},
  2026.
\newblock Accessed 6 October 2026.

\bibitem[{Public Suffix List contributors}(2026)]{psl}
{Public Suffix List contributors}.
\newblock Public suffix list.
\newblock \url{https://github.com/publicsuffix/list}, at commit f7f7b5f4, 2026.

\bibitem[Steinacker-Olsztyn et~al.(2025)Steinacker-Olsztyn, Gosain, and
  Dao]{steinacker2025misinformation}
Nicolas Steinacker-Olsztyn, Devashish Gosain, and Ha~Dao.
\newblock Is misinformation more open? {A} study of robots.txt gatekeeping on
  the web.
\newblock arXiv preprint arXiv:2510.10315, 2025.

\bibitem[Sun et~al.(2007)Sun, Zhuang, and Giles]{sun2007robots}
Yang Sun, Ziming Zhuang, and C.~Lee Giles.
\newblock A large-scale study of robots.txt.
\newblock In \emph{Proceedings of the 16th International Conference on World
  Wide Web (WWW '07)}, pages 1123--1124, 2007.
\newblock \doi{10.1145/1242572.1242726}.

\bibitem[Wilson(1927)]{wilson1927}
Edwin~B. Wilson.
\newblock Probable inference, the law of succession, and statistical inference.
\newblock \emph{Journal of the American Statistical Association}, 22\penalty0
  (158):\penalty0 209--212, 1927.
\newblock \doi{10.1080/01621459.1927.10502953}.

\bibitem[{yc-oss contributors}(2026)]{ycoss}
{yc-oss contributors}.
\newblock yc-oss/api.
\newblock \url{https://github.com/yc-oss/api}, file \texttt{companies/all.json}
  at commit 92a11048, 2026.

\end{thebibliography}

\end{document}